\documentclass[twoside]{article}
\usepackage{fleqn,espcrc2,fps}
\newcommand{\AmS}{{\protect\the\textfont2
  A\kern-.1667em\lower.5ex\hbox{M}\kern-.125emS}}

\newcommand{\psib}{{\bar\psi}}
\def\lsi{\raise0.3ex\hbox{$<$\kern-0.75em\raise-1.1ex\hbox{$\sim$}}}
\def\gsi{\raise0.3ex\hbox{$>$\kern-0.75em\raise-1.1ex\hbox{$\sim$}}}
\newcommand{\lsim}{\mathop{\lsi}}

\newcommand{\R}{{\kern+.25em\sf{R}\kern-.78em\sf{I} 
  \kern+.78em\kern-.25em}}
\newcommand  {\rf} [1]{(\ref{#1})}
\newcommand  {\beq}[1]{\begin{equation}\label{#1}}
\newcommand  {\eeq}   {\end{equation}}
\newcommand  {\bea}   {\begin{eqnarray}}
\newcommand  {\eea}   {\end{eqnarray}}

\newcommand  {\vev}[1]{\left\langle #1 \right\rangle}

\newcommand  {\tr}    {\mbox{Tr}\;}

\title{Monte Carlo Studies of the Dimensionally Reduced 4d $SU(N)$
Super Yang-Mills Theory}
\author{J.\ Ambj\o rn $^{{\rm a}}$, K.N. Anagnostopoulos $^{{\rm b}}$,
W. Bietenholz $^{{\rm c}}$, T. Hotta $^{{\rm d}}$ and J. Nishimura
\address{ Niels Bohr Institute, 
Blegdamsvej 17, DK-2100 Copenhagen \O, Denmark\\
$^{{\rm b}}$ Dept. of Physics, Univ.
of Crete, P.O. Box 2208, GR-71003 Heraklion, Greece \\
$^{{\rm c}}$ Institut f\"{u}r Physik, Humboldt Universit\"{a}t,
Invalidenstr. 110, D-10115 Germany \\
$^{{\rm d}}$ Institute of Physics, Univ. of Tokyo,
Komaba, Meguro-ku, Tokyo 153-8902, Japan}
\thanks{Talk presented by K.N.A. at the HEP 2000 Annual Workshop of the
Hellenic Society for the Study of High Energy Physics at the
University of Ioannina (http://theory.physics.uoi.gr/hep2000/).}
}
\begin{document}

\begin{abstract}

We simulate a supersymmetric matrix model obtained from dimensional
reduction of 4d SU($N$) super Yang-Mills theory. The model is well
defined for finite $N$ and it is found that the large $N$ limit
obtained by keeping $g^2N$ fixed gives rise to well defined operators
which represent string amplitudes. The space-time structure which
arises dynamically from the eigenvalues of the bosonic matrices is
discussed, as well as the effect of supersymmetry on the dynamical
properties of the model. Eguchi-Kawai equivalence of this model to
ordinary gauge theory does hold within a finite range of scale. We
report on new simulations of the bosonic model for $N$ up to $768$
that confirm this property, which comes as a surprise since no
quenching or twist is introduced.

\vspace*{-7mm}

\end{abstract}

\maketitle

\section{Introduction}

Recent excitement in string theory stems from the fact that known
string theories are thought to be perturbative expansions of an 11
dimensional theory called M-theory. The former are believed to be
related by dualities and once we construct a non--perturbative
definition of one of them, then we can also describe the vacua of any
of the other theories. Two models, the IKKT \cite{IKKT} and BFSS
models \cite{BFSS}, have been proposed as possible definitions of
M--theory. Both models are thought to be closely related
\cite{CDS}. For analytic work in this context, see Ref.\ \cite{th}.
The IKKT model (or IIB matrix model) \cite{IKKT} is a
candidate for a constructive definition of non--perturbative type IIB
string theory. If the model possesses a unique vacuum this should
describe the space-time in which we live. Space--time arises
dynamically in this model and one can in principle predict its
dimensionality and low energy geometry. It has even been argued that
the gauge group and matter content of our world can arise from the
solution of this model \cite{IK}.

The IKKT model is a reduction of the 10d $SU(N)$ super Yang--Mills
theory to a point, i.e. we restrict the path integral to be only over
constant field configurations. The model that we investigate in this talk
is a 4d counterpart of the IKKT model. Although it is much simpler
than the original model we hope to capture the essential dynamical
features of the 10d model through full scale numerical simulations
which will yield information about non--perturbative properties in
$d=4$.  Simulations in four dimensions are possible because the model
does not suffer from the sign problem, unlike its higher dimensional
cousins. Several important issues can be studied in depth like the
well definiteness of the model at finite $N$, the large $N$ limit, the
space--time structure and the role of supersymmetry. We can also
address the important dynamical issue of the equivalence of the matrix
model to the original large--$N$ gauge theory in the sense of Eguchi
and Kawai \cite{EK}.  We report on large scale simulations of the
supersymmetric and the bosonic model -- obtained by omitting the
fermions in the action \cite{pap1}. By using a carefully constructed
hybrid-R algorithm \cite{HybR} the computational effort in the SUSY
case increases only as $N^5$ and we are thus able to simulate systems with
size up to $N=48$. The bosonic model is simulated as in
Ref.\ \cite{HNT} up to $N=768$. The results for such large $N$ have not
been published yet.

\section{The model}

The IIB matrix model is given by 
%the partition function
\begin{eqnarray} \nonumber
{\bf Z} &=& \int dA \ e^{-S_{b}} 
\int d \bar \psi d \psi \ e^{-S_{f}} \\
S_{b} &=& -\frac{1}{4g^{2}} \, {\rm Tr} [A_{\mu},A_{\nu}]^{2}
\nonumber \\
S_{f} &=& -\frac{1}{g^{2}} \, {\rm Tr} ( \bar \psi_{\alpha} 
\Gamma^{\mu}_{\alpha \beta} [A_{\mu},\psi_{\beta}])
\end{eqnarray}
where $A_{\mu},\, \bar \psi_{\alpha},\, \psi_{\alpha}$ ($\mu = 1 \dots
d,\, \alpha =1 \dots 2^{\frac{d}{2}-1}$) are complex, traceless
$N\times N$ matrices.  In our case, $d=4$, the $A_{\mu}$ (only) are
Hermitian and we use $\vec \Gamma = i \vec \sigma, \, \Gamma_{4} = 1
\!\! 1$.

The model has $SO(d)$ rotational symmetry, which is the euclidean
version of the Lorentz invariance of the original model before
reduction. The $SU(N)$ gauge invariance of the non--reduced model
becomes
 \begin{eqnarray}
   A_\mu       & \rightarrow&  V A_\mu V^\dagger\nonumber\\
   \psi_\alpha & \rightarrow& V \psi_\alpha  V^\dagger\, ,
   \quad { V\in SU(N)}\nonumber\\
   \label{01}
   \psib_\alpha& \rightarrow&  V \psib_\alpha V^\dagger\, .
 \end{eqnarray}
The ${\cal N}=1$ supersymmetry of the non--reduced model takes the form
\begin{eqnarray}
\delta^{(1)}A_\mu&=& i{\bar\epsilon}_1\Gamma_\mu\psi\nonumber\\
\delta^{(1)}\psi &=&\frac{i}{2}\Gamma^{\mu\nu}[A_\mu,A_\nu]\epsilon_1
\end{eqnarray}
whereas {\it after reduction} the model acquires a second supersymmetry
\begin{eqnarray}
\delta^{(2)}A_\mu &=& 0\nonumber\\
\label{02}
\delta^{(2)}\psi &=&\epsilon_2\, .
\end{eqnarray}
The supercharges can be combined to ${\tilde Q}_1=Q^{(1)}+Q^{(2)}$ and 
${\tilde Q}_2=i(Q^{(1)}-Q^{(2)})$ which obey the commutation
relation
\begin{equation}
\label{03}
 [{\bar\epsilon}_1{\tilde Q}_i,{\bar\epsilon}_2{\tilde Q}_j]
 =-2{\bar\epsilon_1}\Gamma_\mu\epsilon_2 p_\mu\delta_{ij}\, ,
\end{equation}
where $p_\mu$ is the generator of the transformation $\delta^t
A_\mu=c_\mu{\bf 1}_N$. The latter is a symmetry of the action which
also appears after reduction. Eq. \rf{03} suggests that the
eigenvalues of the bosonic matrices $A_\mu$ can be interpreted as
space--time points. In this context, $p_\mu$ is the space--time
translation operator.

By adopting the above interpretation of the model there are several
reasons to believe that the IKKT model ($d=10$) is related to type IIB
superstrings. First, by using the semiclassical correspondence
\cite{IKKT} one obtains the action of the Green-Schwarz type IIB
superstring in the so--called Schild gauge \cite{SCHILD}. Second,
classical solutions corresponding to D-strings are constructed in
Ref.\ \cite{IKKT}. An arbitrary number of D-strings and anti D-strings
can be described as blocks of matrices which interact via the off
diagonal blocks which can change their number and size. Thus the model
can be interpreted to contain a second quantized theory of
D-strings. Third the authors of Ref.\ \cite{FKKT} have obtained the
string field theory supercharges and Hamiltonian of the type IIB
string in the light cone gauge from the Schwinger--Dyson equations
(which describe joining and splitting of {\it fundamental} strings)
that the Wilson loops obey by using only the ${\cal N}=2$
supersymmetry and scaling arguments.  The low energy physics that one
obtains from the model remains a mystery. The authors in
Ref.\ \cite{IK} have proposed that the space--time metric is encoded
in the density correlations of the eigenvalues and that diffeomorphism
invariance stems from the invariance of the model under permutations
of the eigenvalues. They also suggest that the gauge group is obtained
from the clustering of eigenvalues in clusters of size $n$. Then the
low energy theory acquires $SU(n)$ local space-time gauge symmetry.

The first question about our 4d model is whether it is well-defined as
it stands.  Since the integration domain of $dA$ is non-compact,
divergences are conceivable. However, our results \cite{pap1} confirm
the original results of Ref.\ \cite{welldef} for SUSY --- and they
agree with very recent analytic results for the bosonic case \cite{aw}
--- that this model {\em is} well-defined for large enough $N$; there
is no need to impose an IR cutoff. This implies that the only
parameter $g$ is simply a {\em scale parameter} that the theory
determines dynamically.  It can be absorbed by introducing
dimensionless quantities
\begin{equation} \label{dimless}
X_{\mu} = A_{\mu} / g^{1/2} \ ; \quad
\Psi_{\alpha} = \psi_{\alpha}/g^{3/4} \, .
\end{equation}

\section{Numerical Simulations}

For our simulation we start by integrating out the fermionic variables
which can be done explicitly \cite{pap1}. The result is given by $\det
{\cal M} $, ${\cal M}$ being a $2(N^2-1)$ $\times$ $2(N^2-1)$ complex
matrix which depends on $A_\mu$.  Hence the system we want to simulate
can be written in terms of bosonic variables as 
\beq{Z} Z = \int \mbox{d}
A ~ \mbox{e} ^{-S_b} \det {\cal M} \ .  
\eeq 
A crucial point for the present work is that the determinant \ $\det
{\cal M}$ \ is actually real positive. This is shown explicitly in
Ref.\ \cite{pap1}. Due to this property, we can introduce a $2(N^2-1)$
$\times$ $2(N^2-1)$ Hermitian positive matrix ${\cal D}= {\cal M}^\dag
{\cal M}$, so that \ $\det {\cal M} = \sqrt{\det {\cal D}}$, and the
effective action of the system takes the form
\beq{eff_act} S_{\mbox{\scriptsize eff}} = S_b - \frac{1}{2} \ln \det
{\cal D} \ .  
\eeq 

We apply the Hybrid R algorithm \cite{HybR} to simulate this system.
In the framework of this algorithm, each update of a configuration is
made by solving a Hamiltonian equation for a fixed ``time'' $\tau$.
The algorithm is plagued by a systematic error due to the
discretization of $\tau$ that we used to solve the equation
numerically. Special care is taken so that the systematic error is of
order $\Delta \tau^2$, up to logarithmic corrections \cite{pap1}. 
We performed simulations at
three different values of the time step $\Delta \tau$.  Except in
Fig.\ 2, we find that the results do not depend much on $\Delta \tau$
(below a certain threshold), so we just present the results for the
value $\Delta \tau = 0.002$, which appears to be sufficiently small.
Extra care is taken so that the computational effort increases only as
$N^5$ (in the bosonic case the corresponding effort increases only as
$N^3$). Therefore for the supersymmetric case we were able to obtain
3060, 1508, 1296, 436 configurations for $N=16,24,32,48$ respectively.
For the bosonic case, we used 1000 configurations for each $N$.  The
$N\le 32$ simulations were performed on a linux farm at NBI and the
$N=48$ on the Fujitsu VPP500 at High Energy Accelerator Research
Organization (KEK), the Fujitsu VPP700E at The Institute of Physical
and Chemical Research (RIKEN), and the NEC SX4 at Research Center for
Nuclear Physics (RCNP) of Osaka University supercomputers.

\section{The space structure}

In the IIB matrix model, the space coordinates arise dynamically
from the eigenvalues of the matrices $A_{\mu}$ \cite{IKKT}. In general
the latter cannot be diagonalized simultaneously, which implies
that we deal with a non-classical space.
We measure its uncertainty by
\vspace*{-1mm}
\begin{displaymath}
\Delta^{2} = \frac{1}{N} \Big[ {\rm Tr} (A_{\mu}^{2}) -
\, ^{max}_{U \in SU(N)} \sum_{i} 
\{ (UA_{\mu}U^{\dagger})_{ii} \}^{2} \Big] \ ,
\end{displaymath}
\vspace*{-1mm}
and the ``maximizing'' matrix $U$ is also used for introducing the
coordinates of $N$ points,
\vspace*{-1mm}
\begin{equation}
x_{i, \mu} = (U_{max} A_{\mu} U_{max}^{\dagger})_{ii} 
\qquad (i=1\dots N).
\end{equation}
\vspace*{-1mm}
What we are really interested in is their pairwise separation
$r (x_{i},x_{j}) = \vert x_{i}-x_{j} \vert$, and we show the
distribution $\rho (r)$ in Fig.\ 1.
We observe $\rho \approx 0$ at short distances 
($r/\sqrt{g} \lsim 1.5$), hence a UV cutoff is
generated {\em dynamically}. We also see that increasing $N$
favors larger values of $r$. To quantify this effect we measure
the ``extent of space''
\vspace*{-1mm}
\begin{equation}
R_{new} = \int_{0}^{\infty} r \rho(r) \ dr \ .
\end{equation}
\vspace*{-1mm}
Fig.\ 2 shows $R_{new}$ and $\Delta$ as functions of $N$
(at $g=1$). $R_{new}$ is finite in contrast to the quantity
$R=\sqrt{\vev{\frac{1}{N}\tr(A_\mu^2)}}
  \sim\sqrt{\int_o^\infty dr\;r^2\;\rho(r)}$ which diverges
logarithmically as $\Delta\tau\to 0$. This is consistent with 
the prediction by
Ref.\ \cite{welldef},  $\rho(r)\sim r^{-3}$ for large $r$.
The inclusion of fermions enhances $R_{new}$
and suppresses $\Delta$, keeping their product approximately
constant and $\sim g N^{1/2}$. 
We will see that the latter product remains finite in the large $N$
scaling limit. It is a kind of uncertainty principle
for space--time fluctuations. 
The lines in Fig.\ 2 show that both quantities follow the same
power law, $R_{new}, \ \Delta \propto N^{1/4}$, in SUSY 
and in the bosonic case. In particular in the bosonic case 
$R_{new}=1.56(1)g^{1/2}N^{1/4}$,
$\Delta =0.907(3)g^{1/2}N^{1/4}$ so that 
$\Delta^{1/2}\approx 0.58 R_{new}$. In the SUSY case 
$R_{new}=3.30(1)g^{1/2}N^{1/4}$,
$\Delta =0.730(3)g^{1/2}N^{1/4}$ so that 
$\Delta^{1/2}\approx 0.22 R_{new}$.
In SUSY this behavior is 
consistent with the branched polymer picture: there
one would relate the number of points as $N \sim
(R_{new}/\ell )^{d_{H}}$, where $\ell$ is some minimal bond,
which corresponds to the above UV cutoff.
The Hausdorff dimension $d_{H}=4$ then reveals consistency
with our result. In the bosonic case the (same) exponent has a
qualitatively different explanation. It originates from a logarithmic
attractive potential between the eigenvalues of the matrices found in
the one loop approximation of the model \cite{HNT}.
\begin{figure}[hbt]
\label{f01}
\vspace{-5mm}
\def\fpsangle{0}
\epsfxsize=65mm
\fpsbox{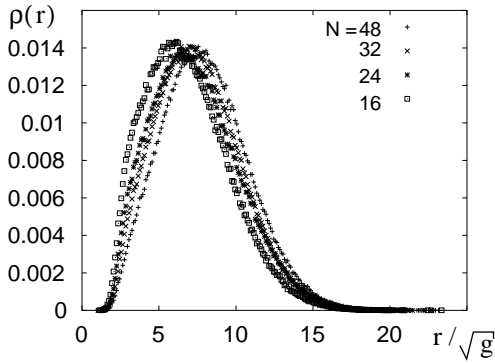}
\vspace{-10mm}
\caption{\it{The distribution of distances between 
space-points in the SUSY case at various $N$.}}
\vspace{-7mm}
\end{figure}
\begin{figure}[hbt]
\label{f02}
\vspace{-8mm}
\def\fpsangle{270}
\epsfxsize=43mm
\fpsbox{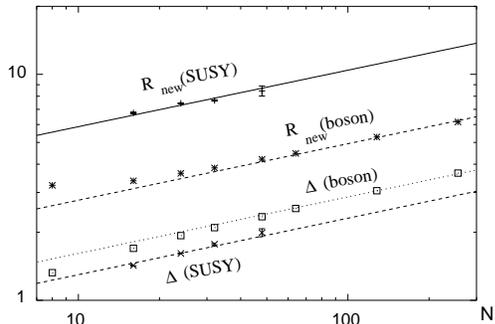}
\vspace{-11mm}
\caption{\it{The ``extent of space'' $R_{new}$ and the
space uncertainty $\Delta$ as functions of $N$ at $g=1$.}}
\vspace{-6mm}
\end{figure}

%Note that $\Delta$ remains finite at large $N$, which agrees
%with superstring theory (for a review, see Ref.\ \cite{Yoneya}).

\vspace*{-1mm}
\section{Polyakov and Wilson loops}
\vspace*{-1mm}

We define the Polyakov loop $P$
%as a Fourier transform of the eigenvalue distribution, 
and the Wilson loop $W$ --- which is conjectured to correspond to
the string creation operator --- as
\vspace*{-1mm}
\begin{eqnarray}
P(p) &=& \frac{1}{N} {\rm Tr} \Big( e^{i p A_{1}} \Big) , \\
W(p) &=& \frac{1}{N} {\rm Tr} \Big( e^{i p A_{1}}e^{i p A_{2}}e^{-i p A_{1}}
e^{-i p A_{2}} \Big) .
\vspace*{-1mm}
\end{eqnarray}
Of course the choice of the components of $A_{\mu}$ is irrelevant,
and the parameter $p \in \R$ can be considered as a ``momentum''.

Now $g(N)$ has to be tuned so that $\langle P \rangle ,\,
\langle W\rangle $ remain finite as $N\to \infty$. 
This is achieved by
\begin{equation} \label{g2N}
g \propto 1/\sqrt{N} \ ,
%\frac{1}{\sqrt{N}} \ ,
\end{equation}
which leads to a beautiful large $N$ scaling; Fig.\ 3 shows the invariance 
of $\langle P \rangle$ for $N=16 \dots 48$ in
SUSY. Also the bosonic case scales accurately \cite{pap1}.
\begin{figure}[hbt]
\vspace{-7mm}
\def\fpsangle{270}
\epsfxsize=50mm
\fpsbox{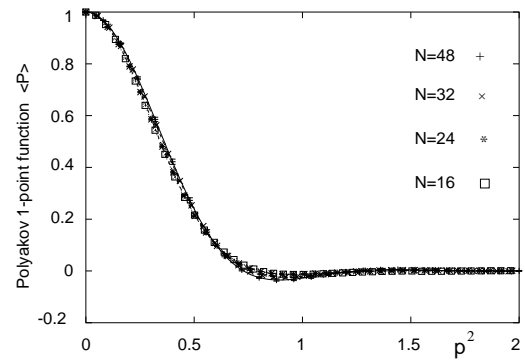}
\vspace{-10mm}
\caption{\it{The Polyakov function in the SUSY case
for various values of $N$ and $g^{2}N=const.$}}
\vspace{-6mm}
\end{figure}

The historic 2d Eguchi-Kawai model \cite{EK} had a 
$(Z\!\!\! Z_{N})^{2}$ symmetry, which implied
$\langle P(p \neq 0) \rangle = 0$, a property which was crucial
for the proof of the Eguchi-Kawai equivalence to gauge
theory. As we see, this property is not fulfilled here, but 
$\langle P(p) \rangle$ falls off rapidly, towards a regime where
the assumption of this proof holds approximately.

We proceed to a more explicit test of Eguchi-Kawai equivalence by
checking the area law for $\langle W(p) \rangle$.  Fig.\ 4 shows that
the area law seems to hold in a finite range of scale for the model
with supersymmetry. Remarkably, the behavior is very similar
\cite{pap1} in the bosonic case\footnote{Recently the area law
behavior was also observed in the 10d bosonic case
\protect\cite{HE}.}. There we further investigated the behavior at
much larger $N$ \cite{prep}, and we observed that the power law regime
does neither shrink to zero --- as it was generally expected --- nor
extend to infinity --- a scenario which seems possible from Fig.\ 4.
At least in the bosonic case its range remains finite at large $N$ as
can be seen in Fig.\ 5.

\begin{figure}[hbt]
%\vspace{-11mm}
\def\fpsangle{270}
\epsfxsize=50mm
\fpsbox{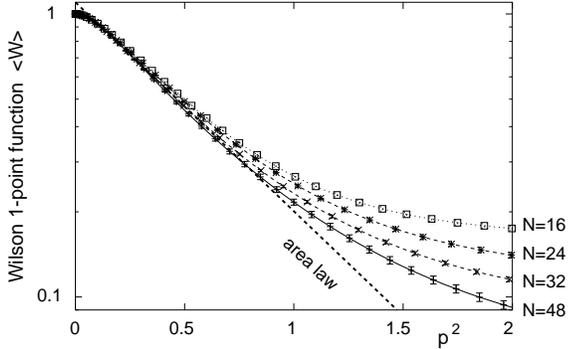}
\vspace{-11mm}
\caption{\it{The Wilson loop in the SUSY case
for various values of $N$ and $g^{2}N = const.$}}
\vspace{-9mm}
\end{figure}

\begin{figure}[hbt]
%\vspace{-11mm}
\def\fpsangle{270}
\epsfxsize=50mm
\fpsbox{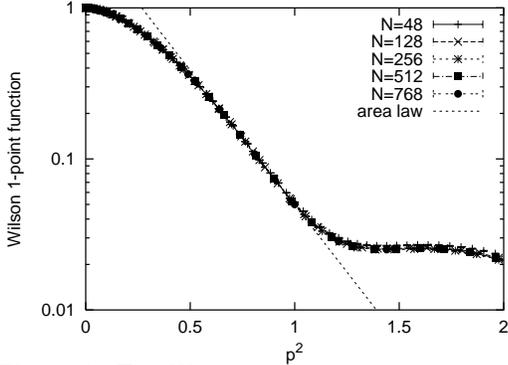}
\vspace{-11mm}
\caption{\it{The Wilson loop in the bosonic case
for various values of $N$ and $g^{2}N = const.$}}
\vspace{-6mm}
\end{figure}

\section{Multipoint functions}
\vspace*{-1mm}

We now consider connected multipoint functions
$\langle {\cal O}_{1} {\cal O}_{2} \dots {\cal O}_{n} \rangle_{con}$,
${\cal O}_{i}$ being a Polyakov or a Wilson loop.
We wonder if it is possible to renormalize all of those
multipoint functions simply by inserting ${\cal O}_{i}^{(ren)}
= Z {\cal O}_{i}$, so that a single factor $Z$ renders all
functions $\langle {\cal O}_{1}^{(ren)} {\cal O}_{2}^{(ren)} 
\dots {\cal O}_{n}^{(ren)} \rangle_{con}$ (simultaneously)
finite at large $N$.

It turns out that such a universal renormalization factor seems 
to exist in SUSY. We have to set again 
$g \propto 1/\sqrt{N}$, and then $Z \propto N$ provides
large $N$ scaling, as we observed for a set of 2, 3 and 4-point 
functions. Two examples are shown in Fig.\ 6. 
Our observation can be summarized by
the SUSY rule
\begin{displaymath}
\vspace*{-1mm}
\langle {\cal O} \rangle = O(1) \ , \quad
\langle {\cal O}_{1} \dots {\cal O}_{n} \rangle = O(N^{-n}) \quad
(n \geq 2).
\end{displaymath}
This implies that large $N$ factorization holds,
$\langle {\cal O}_{1} \dots {\cal O}_{n} \rangle = 
\langle {\cal O}_{1} \rangle \dots \langle {\cal O}_{n} \rangle
+ O(N^{-2})$, as in gauge theory, although
coupling expansions are not applicable here.
\begin{figure}[hbt]
%\vspace{-9mm}
\def\fpsangle{270}
\epsfxsize=50mm
\fpsbox{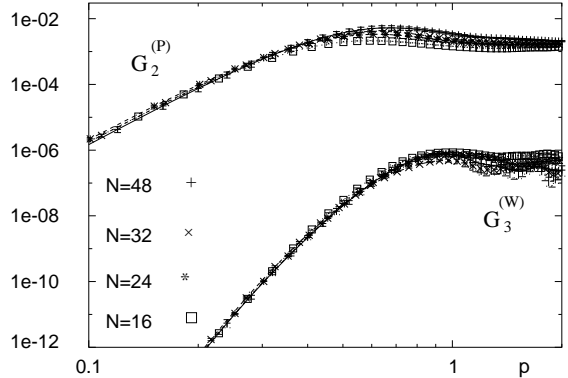}
%\vspace{-5mm}
%\def\fpsangle{270}
%\epsfxsize=40mm
%\fpsbox{wil3.susy.eps}
\vspace{-10mm}
\caption{\it{A Polyakov 2-point function
($G_{2}^{(P)}$ $= \langle [ {\rm Im} P(k)]^{2} \rangle$) 
and a Wilson 3-point function
($G_{3}^{(W)} = \langle [{\rm Im} W(k)]^{2} {\rm Re} W(k) \rangle
- \langle [{\rm Im} W(k)]^{2}\rangle \, 
\langle {\rm Re} W(k) \rangle$)
(both connected), with $g^{2}N=const.$ and renormalization factor
$Z\propto N$, which leads to large $N$ scaling in the SUSY case.}}
\vspace{-5mm}
\end{figure}

For the bosonic case, a $1/d$ expansion \cite{HNT} suggests large $N$
factorization to hold as well, but it also predicts $\langle {\cal
O}_{1} \dots {\cal O}_{n} \rangle = O(N^{-2(n-1)}) \quad (n \geq
2)$. This is confirmed numerically \cite{pap1}: in particular the
3-point functions now require $Z^{3}\propto N^{4}$. Therefore no
universal renormalization factor $Z$ exists in the bosonic case, which
is an important qualitative difference from the SUSY case.

\vspace*{-1mm}
\section{Conclusions}
\vspace*{-1mm}

We reported results from numerical simulations of the 4d IIB matrix
model, both, SUSY and bosonic.  In the SUSY case we varied $N$ up to
48, which turned out to be sufficient to study the large $N$ dynamics.

We confirmed that the model is well-defined as it stands, hence $g$ is
a pure scale parameter.  The space--time
coordinates arise from eigenvalues of the bosonic matrices
$A_{\mu}$. The extent of space--time follows a power law in $N$ with
power of $1/4$. In SUSY this agrees with the branched polymer picture.
Fermions leave the power unchanged but reduce the space--time uncertainty
--- though it remains finite at large $N$. Space--time is quantum with
the uncertainty in determining space--time points to scale together
with the extent of space--time, surviving thus the large $N$ limit.

The large $N$ scaling of Polyakov and Wilson loops and
their correlators requires $g \propto 1/\sqrt{N}$ in SUSY
and in the bosonic case, but the wave function renormalization
is qualitatively different: only in SUSY a universal renormalization
exists. Using a rough argument presented in Ref.\ \cite{pap1} this
suggests that supersymmetry renders the world sheet smoother than in
the bosonic case. Indeed such a phenomenon has been observed
in the dynamical triangulation approach \cite{BBPT}.

The area law for Wilson loops holds in a finite range of scale for the
SUSY and the bosonic case. The latter comes as a surprise, and we
checked up to $N=768$ that this range remains indeed finite. Hence
Eguchi-Kawai equivalence to ordinary gauge theory \cite{EK}, even {\em
without} quenching or twist, may hold in some regime.

\end{document}